\documentstyle[12pt]{article}

\def\pt{$p_T$}
\def\dis{distribution}
\def\az{azimuthal}
\def\dd{$\Delta\phi$}
\def\mm{momentum}

\begin{document} 
\begin{center}  {\Large {\bf Away-side azimuthal distribution in a \\Markovian parton scattering model}}
\vskip .75cm
  {\bf Charles B.\ Chiu$^1$ and  Rudolph C. Hwa$^{2}$}
\vskip.5cm

  {$^1$Center for Particle Physics and Department of Physics\\
University of Texas at Austin, Austin, TX 78712-0264, USA\\
\bigskip
$^2$Institute of Theoretical Science and Department of
Physics\\ University of Oregon, Eugene, OR 97403-5203, USA}
\end{center}

\begin{abstract}
An event generator is constructed on the basis of a model of multiple scattering of partons so that the trajectory of a parton traversing a dense and expanding medium can be tracked. The parameters in the code are adjusted to fit the \dd\ \az\ \dis\ on the far side when the trigger \mm\ is in the non-perturbative region, \pt(trigger) $<$ 4 GeV/c. The dip-bump structure for $1<p_T({\rm assoc})<2.5$ GeV/c is reproduced by averaging over the exit tracks of deflected jets. An essential characteristic of the model, called Markovian Parton Scattering (MPS) model, is that the scattering angle is randomly selected in the forward cone at every step of a trajectory that is divided into many discrete steps in a semi-classical approximation of the non-perturbative scattering process. Energy loss to the medium is converted to thermal partons which hadronize by recombination to give rise to the pedestal under the bumps. When extended to high trigger momentum with \pt(trigger) $>$ 8 GeV/c, the model reproduces the single-peak structure observed by STAR without invoking any new dynamical mechanism.
\end{abstract}

\section{Introduction}

The study of jet quenching in heavy-ion collisions has in recent years become a major area of activity \cite{jw}. Beyond the measurement of the nuclear modification factor  at high \pt, the correlation between jets is particularly effective in revealing the effect of dense medium on the propagation of high-momentum partons through the medium \cite{ka,ja,ca,ja2,ssa}. It has been found that the azimuthal distribution on the away side in the transverse plane opposite a trigger shows structure that depends significantly on the trigger momentum \cite{ja3,jj}. We present in this paper a model based on multiple scattering of partons that can reproduce the major features of that \dis.

The \az\ (\dd) \dis\ that has stimulated a great deal of attention is the one whose trigger momentum is in the range $2.5<p_T<4.0$ GeV/c and associated particles in $1<p_T<2.5$ GeV/c \cite{ssa}. The dip-bump structure that is centered at $\Delta\phi=\pi$ has led to various suggestions on the possible origin of that structure, e.g., Mach cone \cite{jc}, Cherenkov radiation \cite{id,vk}, color wake \cite{jr} and jet quenching in pQCD \cite{iv}. While all such approaches may each contain some element of reality, it seems that a balanced overview can be achieved if the collective response of the system is contrasted by an opposite approach in which a hard parton undergoes multiple scattering with successive Markovian angular deflections. We do not imply that the various possibilities are mutually exclusive. Data will eventually determine the relative importance of the collective behavior of the medium versus the Markovian deflections of partons.
We investigate here the latter approach in a model that is conceptually semi-classical. Since the energies of the partons under consideration are low enough so that pQCD is not reliable, it would be difficult to implement quantum interference in our process of successive collisions as a parton propagates through the dense, but expanding, medium. However, a number of ingredients in the model are adapted from studies based on pQCD and relativistic hydrodynamics, such as fractional energy loss and medium expansion. Thus our approach is not totally orthogonal to other approaches.

We shall construct the model in a Monte Carlo code that will be referred to as Markovian Parton Scattering (MPS) model.
Free parameters in the model are adjusted to reproduce the data on the \dd\ distribution at moderate \pt(trig) and low \pt(assoc). We then calculate the \dis\ at higher \pt\ to show reasonable agreement with the corresponding data. We shall simulate events with varying initial point of hard scattering and separate them into two sets: one, the events in which the recoil parton is absorbed in the medium, and the other in which it emerges from the medium. The absorbed partons are thermalized and hadronize as enhanced background of identifiable away-side jets, while the emerging partons hadronize either by recombination with the thermal partons or by fragmentation to form the away-side jet. The main characteristic of this MPS model is that there is at most only one jet on the away side in each event (sometimes referred to as a deflected jet) in contrast to that in a model, such as Mach cone, for which two symmetrically separated jets are expected on the two sides of the cone tip in each event.  Experimental efforts are currently being made to measure the three-particle correlation \cite{jgu,na}, so that in addition to the trigger particle the correlation between the two particles on the away side can reveal the properties of the event structure.  Our model therefore addresses one of the dynamical possibilities that can contribute to that event structure. When the model is applied to higher trigger momentum ($>$ 8 GeV/c), the two bumps coalesce to a single peak. Sample tracks can be generated to reveal how the nature of the parton trajectories change as the initial hard parton momentum is increased.

\section{The MPS model}

The point of view we adopt in constructing our model is that the trajectory of a parton in the dense medium is not necessarily straight, especially when the initial momentum is not large and the medium density is high in the center.  If the trajectory is curved, then any sensible approach to such a trajectory would be to study it in incremental discrete parts of straight sections.  But in so doing we would have to abandon the coherent response of the medium and focus mainly on the local scattering properties at each step, a choice that is well suited to taking into account the decrease of medium density during the time interval of the parton's passage of the expanding medium.  Our emphasis then puts the model in a semi-classical approximation to the problem, which is basically non-perturbative, since we aim to understand the event structure where the \pt\ range of the associated particles reaches as low as 1 $< p_T ({\rm assoc}) < 2.5$ GeV/c.

To implement MPS, we adopt the  feature that the scattering angle at each time step does not retain any memory of the previous steps.  That is, at each step the scattering angle is randomly chosen to be in the forward cone according to a Gaussian distribution whose width depends on the parton energy and local medium density.  Thus the trajectory of the parton is not only not linear, it is not even smoothly curved.  It can zig-zag through the medium and may not even emerge from it.  We do not regard any individual trajectory to be necessarily realistic; however, just as with path integrals we regard the sum over all such trajectories to represent some meaningful property of the problem, which incorporates a number of reasonable features, such as energy loss and hydrodynamical expansion.   For the nature of energy loss per unit length we adopt the form found in pQCD \cite{bd}, but allow the parameters involved to be varied to fit the data in certain domain.  Such an approach leads us also to a formulation of the changes in the step size due to energy loss in our iterative procedure of determining the complete trajectory of each parton in any event. 
When the parton energy is reduced below a threshold value before reaching the boundary, we regard the parton to be absorbed in the medium.

We now describe the algorithm of the event generator MPS.

\noindent  {\it 1. Scattering angle}

At each scattering point the scattering angle $\alpha$ is randomly chosen from the forward cone defined by a Gaussian \dis
\begin{eqnarray}
G(\alpha,\sigma)={1\over \sqrt {2\pi}\sigma}\exp (-\alpha^2/2\sigma^2),   \label{1}
\end{eqnarray}
where the width $\sigma$ should decrease with decreasing density, but increase with decreasing parton energy. Specifically, we adopt the form
\begin{eqnarray}
\sigma_i=\sigma_s{\rho_i/\rho_1 \over E_i/E_s} \ ,   \label{2}
\end{eqnarray}
where $\sigma_s$ is the   width  at some reference energy $E_s$ to be specified below.  The index $i$ labels  various points on a trajectory with $i=1$ being for the initial hard scattering point. $\rho_i$ and $E_i$ are the density and parton energy at the $i$th point. Since the first step of the recoil parton is directed opposite to the hard parton that produces the trigger particle, there is no angular ambiguity in the first step of the trajectory, so $\sigma_1=0$. Equation (\ref{2}) applies to all subsequent scattering points at $i=2,3,\cdots$. The angle $\alpha$ has at each step no memory of the value of $\alpha$ at the previous step; hence, the process is Markovian.

\noindent  {\it 2. Step size}

How a parton trajectory can be unambiguously divided into discrete steps is not known, so we only introduce density and energy dependencies relative to some scale $\Delta_s$ to be determined phenomenologically. We adopt the form
\begin{eqnarray}
\Delta_i=\Delta_s e^{\rho_1-\rho_i} \sqrt{E_i/E_s} \ ,   \label{3}
\end{eqnarray}
where the $\sqrt E$ dependence is an adaptation from the energy-loss behavior obtained in Ref.\ \cite{bd} (BDMPS)
\begin{eqnarray}
{dE\over dx} \propto -\sqrt E\ ,   \label{4}
\end{eqnarray}
which implies that $d\sqrt E/dx$ is approximately constant  at some fixed density. We introduce the density dependence in Eq.\ (\ref{3}) by requiring $\Delta_i$ to decrease with increasing density, but not in a power law  in order to avoid divergence at very low density, since each step will be mapped to a finite time interval during which a parton can only traverse a finite distance even at zero density. $\rho_1$ sets the dimensionless scale of the density dependence of $\Delta_i$.

\noindent  {\it 3. Time steps}

Since the step size $\Delta_i$ does not depend linearly on the step index $i=1,2,\cdots$, indeed, not even monotonically, the time interval $\tau_{i+1}-\tau_i$ is not constant. Since the velocity is essentially constant (at $c$), the time interval is proportional to $\Delta_i$, which we define to be the spatial distance from step $i$ to $i+1$. The proportionality constant is immaterial, since $\tau$ is a dimensionless scaled time variable. Thus we simply set
\begin{eqnarray}
\tau_{i+1}-\tau_i=\Delta_i/\Delta_1 ,  \label{5}
\end{eqnarray}
so that with $i=1$ marking the point of the hard scattering, we have $\tau_1=1$.

\noindent {\it 4. Initial point}

We model only the events where the initial hard scattering occurs near the surface of the dense medium so that one  of the scattered partons goes immediately out of the medium and generates a particle that triggers the event.  We defer to a later discussion the issues related to how that parton hadronizes, and begin tracking the recoil parton with initial energy $E_1$. We consider only the trajectories in the transverse plane with the initial points being randomly distributed in a narrow circular strip just inside the boundary with radius $r$ confined to $\xi R_0<r<R_0$, where $R_0$ is the initial radius, and $\xi$ a number close to 1 to be specified. The azimuthal angle of the strip is bounded by $|\Delta\phi|<\phi_c$, measured from the trigger direction. We shall, for convenience, use the notation $\xi_1=\sin\phi_c$, the value of which will also be specified later.

\noindent {\it 5. Medium expansion}

The radial flow of the medium in a central collision leads to an expansion of the circle in the transverse plane with its radius increasing as
\begin{eqnarray}
R_i=R_0 \exp [H(\tau_i-1)] ,   \label{6}
\end{eqnarray}
where $H=0.07$, given in Ref.\ \cite{pfk} for Hubble-like expansion. We assume, for simplicity, that the radial distance $r$ of all points inside the circle increases in proportion to $R$, and that the density of the medium decreases uniformly with $\tau$ independent of $r$, i.e., 
\begin{eqnarray}
\rho_i(r,\tau_i)={\rho_s\over \tau_i}\Theta(R-r) ,  \label{7}
\end{eqnarray}
where $\Theta$ is the step function. We shall consider only central collisions so that we may ignore the complications arising from elliptic flow for non-central collisions.

\noindent {\it 6. Energy loss}

Having described the geometrical and scattering aspects of the problem, we can now discuss the energy-loss consideration appropriate for our problem. Theoretical studies of that problem have used the nuclear modification factor $R_{AA}$ as a guide to determine certain parameters \cite{kje,ad}, but the data on $R_{AA}$ include all events, not just the ones we consider, viz., the events where hard scattering occurs near the surface. Nevertheless, apart from the values of phenomenologically determined parameters, we adopt the BDMPS form for energy loss \cite{bd}, and rewrite Eq.\ (\ref{4}) more explicitly as 
\begin{eqnarray}
{dE\over dx}=-\kappa_1 \sqrt{E/E_s}     \label{8}
\end{eqnarray}
for some constant $\kappa_1$. Integrating Eq.\ (\ref{8}) over one step from $i$ to $i+1$, we get
\begin{eqnarray}
\sqrt {E_{i+1}}=\sqrt{E_i} - {\kappa_1 \Delta_i\over 2\sqrt{E_s} }.   \label{9}
\end{eqnarray}
Substituting Eq.\ (\ref{3}) into the above, we obtain 
\begin{eqnarray}
E_{i+1} = E_i \ [1-\kappa \exp(\rho_s-\rho_i)]^2 ,    \label{10}
\end{eqnarray}
where
\begin{eqnarray}
\kappa={\kappa_1\Delta_s\over 2 E_s},    \label{11}
\end{eqnarray}
a parameter that replaces the unknown $\kappa_1$, and characterizes the change in the energy of the parton from one step to the next.

\noindent{\it 7. Absorbed partons}

If the initial parton energy is not too high, there is a significant probability that a recoil parton may not emerge from the medium. We shall set the threshold energy for a parton to emerge at 0.3 GeV so that we terminate the tracking of a trajectory when $E_i<0.3$ GeV.  When we consider the data with trigger momentum in the range $2.5<p_T({\rm trigger})<4$ GeV/c, we start all trajectories with $E_1=4.5$ GeV. The energy lost by the recoil parton to the medium during multiple scattering are converted to thermal energy that 
is distributed over a wide range of rapidity because of longitudinal expansion. Thus we include only the effect of the last step since it is closest to the away-side surface. Specifically, if $i=N$ is the last step either before exiting or being absorbed, then we count the deposited energy $E_{\rm dep}$ toward   the \dd\ \dis\ in the following way. If the parton does not emerge, i.e., absorbed, then we let $E_{\rm dep}=E_{N-1}$. If the parton emerges with $E_{\rm exit}$, then $E_{\rm dep}=E_{N-1}-E_{\rm exit}$. We average $E_{\rm dep}$ over all events, and distribute the $\left<E_{\rm dep}\right>$ uniformly over the interval $-1.5<\Delta\phi-\pi<1.5$.
 
\noindent{\it 8. Exit partons and their hadronization}

A parton that emerges from the medium will hadronize by recombination with thermal partons, since the range of \pt\ for the associated particles is $1-2.5$ GeV/c \cite{ssa}. For simplicity, we ignore proton production (although the $p$ to $\pi$ ratio may not be small) and assume the average energy that the thermal partons contribute is 0.3 GeV. Thus only the exit partons with momenta in the range $0.7-2.2$ GeV/c will be collected to determine the $\Delta\phi$ \dis\ of the associated particles. The multiplicity \dis\ of the associated particles will be taken to be a factor 1.7 larger than the exit-parton \dis, since even at such low \pt\ a parton may generate shower partons that can lead to more hadrons by recombination with thermal partons. No hadronization scheme of such non-thermal but moderately soft exit  partons is reliable here. We use the 1.7 factor as  a reasonable ansatz.

\noindent{\it 9. Thermalization of energy lost}

Consider first the thermal partons emitted on the away side without the effect of energy loss by the hard parton. Hadronization of thermal partons has been treated before \cite{hy}; we adopt the same formalism here.
For thermal partons having a \dis\ 
\begin{eqnarray}
{\cal T}(q)=q{dN_q^{\rm th}\over dq}=C_0qe^{-q/T_0} \ ,     \label{12}
\end{eqnarray}
where $q$ is the transverse momentum, and $C_0$ and $T_0$ are parameters to be discussed below.  The recombination of two such partons results in the thermal \dis\ of pions in transverse momentum $p_T$ \cite{hy}
\begin{eqnarray}
H_\pi(p_T)={dN_\pi^{\rm th}\over p_Tdp_T}={C_0^2\over 6}e^{-p_T/T_0}    \label{13}
\end{eqnarray}
per unit rapidity per radian. The transverse energy density in $(dyd\phi)$ is then
\begin{eqnarray}
E_0=C_0^2T_0^3/3 \ .    \label{14}
\end{eqnarray}
With this regarded as the reference energy without the deposit of energy lost by the recoil parton,  the inclusion of that energy in the medium raises the temperature from $T_0$ to $T$. With the assumption that the deposited energy $E_{\rm dep}$ computed above creates a pedestal above the background for the azimuthal angle in the range $-1.5<\Delta\phi-\pi<1.5$ only (since all the recoil partons move toward that range), the modified thermal energy per radian is
\begin{eqnarray}
E_T=E_0+{E_{\rm dep}/ 3}\ ,  \label{15}
\end{eqnarray}
where the  1/3 factor on the right is due to dividing the deposited energy by 3 rad.

The parameters $C$ and $T$  for single-particle \dis\ have been determined previously \cite{hy}, but their values $C=23.2$ GeV$^{-1}$ and $T=0.317$ GeV are appropriate only for soft partons in the immediate neighborhood of hard partons near the surface  of the near side
 for recombination. The hadronization of the deposited energy that we are studying here occurs much later, since the recoil partons that are mostly absorbed have to travel across the whole medium in the transverse plane after the hard scattering, so the corresponding thermal partons on the away side are emitted at least 10 fm/c later. Thus the relevant $C_0$ and $T_0$ that concern us here are much lower. We take them to be about half as much: $C_0=10$ GeV$^{-1}$ and $T_0=0.17$ GeV, the latter being approximately the freeze-out temperature. These parameters refer to the background that is usually subtracted in the data analysis before the \dd\ \dis\ is shown.

\noindent{\it 10. Pedestal in the $\Delta\phi$ \dis}

To translate the enhanced thermal energy to the increase of particle multiplicity in the away-side $\Delta\phi$ \dis, let us start with the modified thermal \dis, which we require to have the same form as in Eq.\ (\ref{13}) but with $T_0$ replaced by $T$, and $C_0=C$ remaining unchanged. Thus, as in Eq.\ (\ref{14}), the enhanced thermal energy per radian is 
\begin{eqnarray}
E_T=C^2T^3/3.
\label{16}
\end{eqnarray}
Using Eq.\ (\ref{15}) we can determine $T$. For the multiplicity we only need to integrate $H_\pi(p_T,T)$ between $p_1$ and $p_2$, getting
\begin{eqnarray}
N_{12}(T)=\int_{p_1}^{p_2} dp_T\,p_T H_\pi(p_T, T)={C^2T^2\over 6}G_{12}(T) \ ,  \label{17}
\end{eqnarray}
where
\begin{eqnarray}
G_{12}(T)=(1+x_1)e^{-x_1}-(1+x_2)e^{-x_2}    \label{18}
\end{eqnarray}
with $x_i=p_i/T$. The height of the pedestal under the peaks of the exit tracks is the difference between $N_{12}(T)$ and the background at $T_0$, i.e.,
\begin{eqnarray}
\left.
{dN\over d\Delta\phi}\right|_{ped}=N_{12}(T)-N_{12}(T_0).  \label{19}
\end{eqnarray}
This way of determining the pedestal height has been considered before for the near-side \dis\ \cite{ch}.

This completes our description of the algorithm to generate events with jet correlation. Model parameters used   to fit the data of Ref.\ \cite{ssa} are as follows:

(a)  {\it Geometry and expansion:}

\indent\indent $R_0=6$\,f,\ $\xi=\xi_1=0.8$,\ $H=0.07$

(b)  {\it Initial conditions: }

\indent\indent (b$_1$) $\rho_1=0.63$,   

\indent\indent (b$_2$) $\sigma_2=\sigma_s/2=0.88,\ \Delta_1=\Delta_s=1.9\,{\rm f}$   at the reference energy $E_s=5$ GeV. 

(c)  {\it Energy loss:}

\indent\indent $\kappa=0.17$

The parameters $\rho_1, \sigma_s, \Delta_s$ and $\kappa$ are free to adjust, while the others are taken to be at the default values indicated above.

\section{Results of simulation}

In Fig.\ 1 we show some sample trajectories of recoil partons that either (a) emerge from the medium, or (b) are absorbed. There are 15 samples shown in each category. Each trajectory moving initially along the $Y$ direction is the recoil of a hard parton that is pointed in the $-Y$ direction. All starting points are located in a narrow strip just inside the smallest circle of radius $R_0$. The two larger circles are the boundaries of the medium at later times, the larger one of which does not represent the final size. It is clear in Fig.\ 1(a) that the exit tracks all start from the two sides and bend successively in the same direction to emerge with short paths in the medium. They hadronize in the region where $|\Delta\phi|$ is about 1 rad from $\pi$.
On the other hand, Fig.\ 1(b) shows that the absorbed tracks all start near the center of the strip with small $|X|$ component and bend in either direction randomly so that the path lengths in the medium are long and get absorbed before reaching the final boundary (not shown). Each track by itself may not be realistic due to our discrete multiple-scattering approximation; however, when averaged over all events, we get three types, one a left-bending jet, another a right-bending jet, and finally thermalized tracks that end in the medium with a wide range of $\Delta\phi$ \dis\ in their terminal points. We regard these three types of trajectories to be quite representative of the realistic situation where the observed $\Delta\phi$ \dis\ on the away side has the double-bump structure. The most important characteristic of this MPS model is that an exit jet can only be either a left-bending or a right-bending one. There can never be two simultaneous jets on the away side in any one event in contrast to a scenario that a model like the Mach cone describes. The two types of models cannot be distinguished by di-hadron correlation, but can possibly be identified in tri-hadron correlation.

In Fig.\ 2 we show the $\Delta\phi$ \dis\ of our simulation using the parameters listed at the end of the preceding section. On the horizontal axis we use the $\varphi$ variable to denote $\Delta\phi-\pi$ so that $\varphi=0$ corresponds to the direct opposite of the trigger direction. The dashed line shows the pedestal arising from the thermalization of the energy deposited in the medium in the last step, distributed uniformly in the region $|\varphi|<1.5$. 
From the average energy lost by the partons at the last step  before exiting or being absorbed, we obtain the enhanced temperature $T$ at 0.2 GeV/c, an increase of 0.03 GeV/c over the background $T_0$. The height of the pedestal is $dN/d\varphi=0.0165$.
What stands between the dashed line and the solid line is the average hadron multiplicity $dN/d\varphi$ due to exit tracks. 
A factor of 1.7 has been included in the ratio of hadron to exit-parton multiplicities.
The  data in Fig.\ 2 are from PHENIX for $2.5<p_T({\rm trigger})<4$ GeV/c and $1.0<p_T({\rm assoc})<2.5$ GeV/c  \cite{ssa}. 
Given that the data are in a region that is highly non-perturbative and involves many aspects of physics that are complicated, the agreement between the model output and the data is quite acceptable.

The model that we have constructed is aimed at describing hadron correlation in the trigger momentum range less than 4 GeV/c.  Now, with the geometrical and dynamical parameters fixed, we can simulate events at higher trigger momentum and compare with existent data from STAR. 

\section{Higher Trigger Momentum}

STAR has data on correlation with trigger momentum in the range $8<p_T({\rm trigger})<15$ GeV/c, and associated particles in the ranges $4<p_T({\rm assoc})<6$ GeV/c and $p_T({\rm assoc})>6$ GeV/c \cite{ja3}. In these very wide ranges of \pt\ there arises new issues that we must consider. Firstly, the trigger momentum is not only high, but is uncertain up to a factor of 2. For a hadron with $p_T>8$ GeV/c the hadronization process is dominated by fragmentation \cite{hy}, but even if \pt(trigger) is fixed, the fragmenting hard parton can have a large range of possible momentum $k$. Now, with \pt(trigger) varying from 8 to 15 GeV/c, the hard parton momentum $k$ has an even larger degree of uncertainty.

Consider next the away side. The associated particle momentum can vary from 4 GeV/c to the trigger momentum. Such a scenario introduces two biases. The first bias is well known. At high trigger momentum, the hard scattering point is likely to be near the surface of the trigger side. We have already incorporated such a bias in our simulation model. The second bias is less familiar. The associated-particle momentum range on the away side selects preferential hard-parton momentum, especially when the latter has a wide range of uncertainty. The reasoning is as follows. First of all, hadronization on the away side, when \pt(assoc)$>$ 4 GeV/c, is dominantly fragmentation instead of recombination because the thermal parton \dis\ in Eq.\ (\ref{12}) is significantly reduced in both $C_0$ and $T_0$. The demarkation between recombination and fragmentation found to be at $p_T\sim 8$ GeV/c \cite{hy} applies only to the near side where the thermal \dis\ is strong. Now, on the away side at much later time the demarkation boundary should be shifted downward substantially, and 4 GeV/c is reasonable. An exit parton on the far side, initiated by a hard scattering on the near side that gives rise to a trigger momentum $>8$ GeV/c, must fragment to produce the detected associated particle. The fragmentation function allows a wide range of possible exit parton momentum, which in turn allows  a wide range of hard parton momentum $k$. Clearly, the lower range $4<p_T({\rm assoc})<6$ GeV/c is biased toward lower $k$, while the higher range \pt(assoc)$>$6 GeV/c requires higher $k$. The consequence is that in this section on higher trigger momentum we cannot fix $k$ as we have done in Sec. II.

Since the associated particles in the latter range are dominated by those with \pt\ near the lower limit, we do not expect the $k$ values for the two ranges to differ by too much. We shall take them to be 10\% apart. Specifically, we set $k_1=14.5$ GeV/c and $k_2=16$ GeV/c, respectively, for the two regions. In Fig.\ 3 we show the results of our simulation for the two cases when all other parameters are held fixed as in Sec.\ II. We take the background as given by the data without attempting to simulate the semi-hard processes underlying the peak that we associate with the exit tracks. Fragmentation of valence quarks is considered for both (a) and (b) and the results are shown by the solid lines. The agreement with data is excellent. There are no absorbed tracks. All recoil partons punch through the medium and give rise to the peaks around $\varphi=0$. MPS is capable of describing correctly the height and widths of the peaks in both intervals of the momenta of the associated particles.

\section {Conclusion}

The MPS model that we have investigated provides a satisfactory description of the partonic process in a dense and expanding medium, and therefore offers an alternative explanation of the dip-bump structure in the \dd\ \dis\ that seems to suggest a collective response of the medium to the recoil  parton.  Although the possibility that both mechanisms being coexistent in reality cannot be excluded at this point, we have tacitly assumed that the collective response is negligible.
Since our model can also explain the observed behavior when the trigger momentum is high, it unifies the double-bump and single-peak phenomena without invoking different physical mechanisms.

By varying different adjustable parameters in the model to fit the data at  various momentum ranges, we have learned a number of properties of the system. The deflected jets that emerge on the sides of the medium in the transverse plane, when \pt(trigger) is $<$ 4 GeV/c, have large scattering angles without losing too much momenta, which are inferred from $\sigma_s=0.88$ and $\kappa=0.17$. But as \pt(trigger) is increased, the probability of deflection is rapidly reduced, and the trajectories become mainly linear and go straight through the medium, with the number of absorbed tracks also reduced. These are physically reasonable characteristics of hard partons traversing a medium. What surprises us is the sensitivity of the results to the hadronization mechanism applied to the production of hadrons on the far side.

From previous investigation of single-particle \dis s it is known that thermal-shower recombination dominates over shower-shower recombination (i.e., fragmentation) for $p_T<8$ GeV/c \cite{hy}. On the basis of that we have expected the same to occur for hadronization on the away side. However, we could not reproduce the data in Fig.\ 3 until we realized that when the recoil partons reach the far side the time would be greater than 10 fm/c after the hard scattering, and the thermal environment of an exit parton should be very different from the scenario on the near side where the single-particle spectra at high \pt\ are more relevant. Thus with thermal-shower recombination suppressed the predicted height of the \dd\ \dis\ in Fig.\ 3 (a) based on fragmentation can be brought closer to the data. Another unexpected property is the bias related to the associated-particle cut. When that cut is increased, we had to increase the recoil parton momentum to raise the predicted height of the peak in Fig.\ 3 (b). A 10\% increase results in a good agreement with the data.

Although many parameters have been used in the model, as is usual in  event generators, we have found the investigation to be very satisfactory, since it has revealed to us various features of the complicated process that we could not have learned otherwise. As a result of this study, we can make one prediction that departs from our previous expectation. Since fragmentation dominates the hadronization process on the away side, we predict that the proton-to-pion ratio, $R_{p/\pi}$, among the associated particles should be small, even at $p_T\sim 4$ GeV/c. This  is a drastic departure from the situation on the near side where the dominance of recombination would imply larger $R_{p/\pi}$. Verification of this prediction would give a strong support to the study described here.

\section*{Acknowledgment}
  This work was supported, in part,  by the U.\ S.\ Department of Energy under
Grant No. DE-FG02-96ER40972.

\newpage
\begin{center}
\section*{Figure Captions}
\end{center}

\begin{description}
\item
Fig.\ 1. (color online) Samlple tracks of recoil partons that either (a) leave the medium, or (b) are absorbed. The circles represent the cross section of the medium at three different times, the innermost one being at the initial time, the outermost one not at the final time of the expansion.

\item
Fig.\ 2. (color online) $\Delta\phi$ distribution on the away side. Data are from \cite{ssa} for $2.5<p_T^{\rm trig}<4$ GeV/c and $1<p_T^{\rm assoc}<2.5$ GeV/c. The dashed line is the pedestal calculated from energy deposited in the medium in the last step, and the solid line includes the contribution from the exit tracks that give rise to the deflected jets.

\item
Fig.\ 3. (color online) $\Delta\phi$ distribution on the away side for $8<p_T^{\rm trig}<15$ GeV/c and (a) $4<p_T^{\rm assoc}<6$ GeV/c and (b) $p_T^{\rm assoc}>6$ GeV/c. Data are from \cite{ja3}, and the lines are from  calculation in MPS assuming fragmentation only, with the dashed horizontal lines being assumed to represent the background.

\end{description}
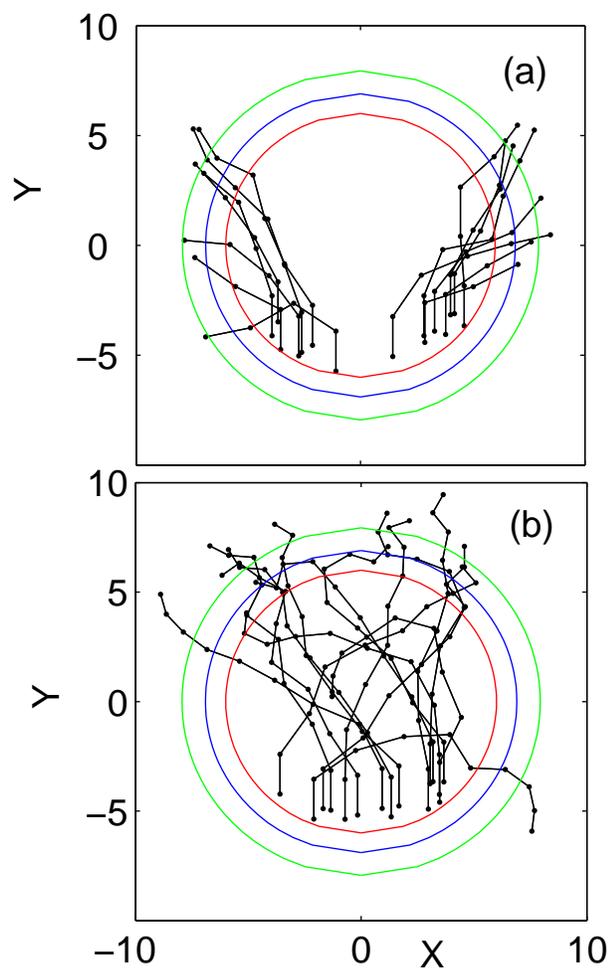
\end{document}